\documentstyle[epsfig]{mn}
\def\bi{\begin{itemize}}
\def\ei{\end{itemize}}
\def\be{\begin{equation}}
\def\ee{\end{equation}}
\def\ba{\begin{eqnarray}}
\def\ea{\end{eqnarray}}
\def\bse{\begin{subequations}}
\def\ese{\end{subequations}}

\title[A Plausible Galactic Spiral Pattern and its Rotation Speed]{A Plausible Galactic Spiral Pattern
and its Rotation Speed}

\author[M. Martos, X. Hernandez, M. Y\'a\~nez, E. Moreno and B. Pichardo]
{M. Martos$^1$, X. Hernandez$^1$, M. Y\'a\~nez$^1$, E. Moreno$^1$, and B. Pichardo$^2$\\
$^1$ Instituto de Astronom\'\i a,
Universidad Nacional Aut\'onoma de M\'exico
A. P. 70--264,  M\'exico 04510 D.F., M\'exico \\
$^2$ University of Wisconsin,
Department of Astronomy
475 N. Charter St.
Madison, WI 53706, U.S.A.\\
}
\date{\today}
\begin{document}
\maketitle
\begin{abstract}

We report calculations of the stellar and gaseous  response to a Milky Way mass distribution model
including a spiral pattern with a locus as traced by K-band observations, over imposed on the axisymmetric
components in the plane of the disk. The stellar study extends calculations from previous work concerning
the self-consistency of the pattern. The stellar response to the imposed spiral mass is studied
via computations of the central family of periodic and nearby orbits
as a function of the pattern rotation speed, $\Omega_p$, among other parameters.              
A fine grid of values of $\Omega_p$ was explored
ranging from 12 to 25 $km~s^{-1}~kpc^{-1}$. Dynamical self-consistency is highly sensitive to $\Omega_p$,
with the best fit appearing at 20 $km~s^{-1}~kpc^{-1}$. We give an account of
recent independent pieces of theoretical and observational work that are  dependent on the   
value of $\Omega_p$, all of which are consistent with the value found here; the recent star formation history
of the Milky Way, local inferences of cosmic ray flux variations and Galactic abundance patterns. 
The gaseous response, which is also a function of $\Omega_p$, was calculated via 2D hydrodynamic simulations
with the ZEUS code.
For $\Omega_p = 20~km~s^{-1}~kpc^{-1}$, the response to a two-armed pattern is
a structured pattern of 4 arms, with bifurcations along the arms and interarm features. The
pattern resembles qualitatively the optical arms observed in our Galaxy and other galaxies.
The complex gaseous pattern
appears to be linked to resonances in stellar orbits. Among these, the 4:1 resonance plays an
important role, as it determines the extent of the stellar spiral pattern in the
self-consistency study here presented. Our findings seemingly confirm predictions by Drimmel
and Spergel (2001), based on K band data.                      

\end{abstract}

\section{INTRODUCTION}

The comparison of near-infrared and optical images of external galaxies reveals interesting differences.
Striking examples are M81 and NGC 2997 (see pictures in Elmegreen, D., 1981; and Block et al 1994,
respectively). It is common to observe a smooth, simple 2-armed K band pattern but a more
complex pattern in the optical blue band, often suggesting more arms and bifurcations (segments of
arms that appear to be connected to a K band arm but are not detectable in the infrared).
A two-armed smooth structure underlying a more complex morphology also appeared in 
the work of Grosb{\o}l, Pompei and Patsis (2002): in a K band study of 53 nearby spirals,
most galaxies displayed a grand-design, two-armed, symmetric pattern in their inner regions which
often breakups into tighter winded, multiple arms further out. Nonlinear effects were invoked
to explain such morphology. 

In recent work, data from COBE-DIRBE have shed light into the Milky Way spiral pattern.
Drimmel (2000), and Drimmel \& Spergel (2001) have presented a comprehensive picture of how
this pattern is like, presenting emission profiles of the Galactic plane in the
K band and at 240 $\mu m$. The former data set, which suffers little absorption and traces density
variation in the old stellar population, is dominated by a two-armed structure with a minimum
pitch angle of 15.5$^\circ$. At 240 $\mu m$, the pattern is consistent with the
standard four-armed model, that corresponding to the distribution of the youngest stellar
populations delineated by HII regions.

The conventional picture of the spiral pattern of our Galaxy maps at least 4 arms, named
Norma, Crux-Scutum, Carina-Sagittarius and Perseus (for a recent review see Vall\'ee 2002, who also
reports a likely pitch angle of 12$^\circ$ for this pattern). Additionally, features such as
the Orion spur at the Solar neighborhood have been revealed (Georgelin \& Georgelin 1976).
Drimmel (2000) laid down, from the comparison, the hypothesis that the 4-armed structure is
the gas response to the 2-armed ``stellar" pattern. 

Assuming that indeed the K band data is by far a better tracer of mass than the optical data
of spiral structure, in this work we model the spiral pattern from the locus and
pitch angle of Drimmel (2000) and study its self-consistency, a requirement we consider it must
satisfy. As the answer strongly depends on the pattern speed, this study should yield a value
for this fundamental Galactic parameter. 

The value of the pattern speed of the Galaxy has been a matter of controversy for a long
time. From the values proposed by Lin, Yuan \& Shu (1969) of $\Omega_p = 11-13~km~s^{-1}~kpc^{-1}$,
numbers in the range of 10-60 $km~s^{-1}~kpc^{-1}$ have been used in the literature. In a previous
paper (Pichardo et al 2003, hereafter P1), we explored the stellar dynamics in a full axisymmetric
model for our Galaxy superimposing our modeling of mass distribution
for the spiral pattern: the locus of Drimmel (2000), the optical locus of Vall\'ee (2002), and a
superposition of both. P1 did not assume the usual simple 
perturbing term (a cosine term for the potential) that had been used in the 
literature in the modeling of spiral arms:    
it is precisely the very prominent spiral structure in red light what
suggests to us that such structure should be considered an important galactic component worthy
of a modeling effort beyond the simple perturbing term. P1 modeling consists of a superposition
of oblate spheroids for the spiral. Two different values of $\Omega_p = 15, 20~km~s^{-1}~kpc^{-1}$,
were tried. The
best self-consistency was achieved with $\Omega_p = 20~km~s^{-1}~kpc^{-1}$ for the locus and pitch
angle of Drimmel (2000). Figure 7 of P1, a mosaic of our self-consistency test -- explained below --
for parameters including the global spiral mass and loci,  exhibits a remarkably good
response for this case which rules out  the other 3 cases in the panel. In fact,
the behavior at $\Omega_p = 20~km~s^{-1}~kpc^{-1}$ is so good that one can hardly envision an
improvement on general theoretical grounds. The question that comes to mind is, given the spiral
locus and pitch angle of Drimmel (2000), our adopted mass distribution modeling, 
and the spiral mass implied by observations of external galaxies applied to those parameters, now
fixed, are there other values of $\Omega_p$ which also satisfy our self-consistency criterion to
that accuracy?
In this paper, we extend the self-consistency calculations to a finer
range of values of $\Omega_p$ in order to answer that question. Once the best value is found, we
put it to the test in two ways: calculating the gaseous response to find out whether it is consistent with 
the observed optical Galactic spiral pattern, which amounts to a test of Drimmel's hypothesis. Finally,
we review recent work using independent data sets which are sensitive to the value of $\Omega_p$. 
This is another way of testing our prediction against "nature", and not only versus
different modeling approaches, as the subjects of those independent studies are quite different from
our Galactic modeling effort.

A continued line of work by Contopolous and collaborators (see, v.g. Patsis,
Grosb{\o}l and Hiotelis 1997 and references therein) has provided the framework to study the response
of gaseous disks to spiral perturbations. In that paper, a comparison between
SPH models with Population I features observed on B images of normal, grand design galaxies,
showed that the 4:1 resonance generates a bifurcation of the arms and interarm features. Furthermore,
Contopoulos \& Grosb{\o}l (1986,1988) had shown that the central family of periodic orbits do not
support a spiral pattern beyond the position of the 4:1 resonance, which thus determines the extent
of the pattern. Weak spirals can extend their pattern up to corotation, from linear theory.
A phenomenological link between resonances, the angular speed, and the stellar and gas patterns
in spirals is complemented  by the study of Grosb{\o}l and Patsis (2001) using deep K band surface
photometry to analyze spiral structure in 12 galaxies. They find that the radial extent of the two-armed pattern 
is consistent with the location of the major resonances: the inner Lindblad resonance (ILR),
the 4:1 resonance, corotation and the outer Lindblad resonance (OLR). For
galaxies with a bar perturbation, the extent of the main spiral was better fitted assuming it is 
limited by corotation and the OLR. Using $H_0 = 75~km~s^{-1}~Mpc^{-1}$, the pattern speed
was found to be for the entire sample of the order of 20 $km~s^{-1}~kpc^{-1}$, and remarkably, not a
sensitive function of morphological type or total mass.

In the following section  we describe our results for 
the stellar orbital response to the imposed spiral pattern, through which $\Omega_{p}$ is determined. 

\section{Orbital Self-Consistency Modeling, Inferring $\Omega_{p}$}   

As in P1 our axisymmetric Galactic model is that of Allen \& Santill\'an (1991), which
includes a bulge and a flattened disk proposed by Miyamoto \& Nagai (1975), together with a
massive spherical dark halo. We coupled to this mass distribution 
a spiral pattern modeled as a superposition of inhomogeneous oblate spheroids along
a locus that fits the K band data of Drimmel (2000), with a pitch angle of 15.5$^\circ$.       
P1 describes in detail the parameters of the spheroids, which briefly are:
the minor axis of the spheroids is perpendicular to the Galactic plane and its length is 0.5 kpc;
the major semi-axes have a length of 1 kpc. Each spheroid has a similar mass distribution.
Different density laws, linear and exponential, were analyzed, finding no important differences.
 
The total mass in the spiral is fixed such that the local ratio of spiral to background (disk)
force have a prescribed value. Seeking sensible values for this ratio, we used the empirical result of 
Patsis, Contopoulos and Grosb{\o}l (1991), where self-consistent models for 12 normal spiral
galaxies are presented, a sample including Sa, Sb and Sc galaxies. Their Figure 15 shows
a correlation between the pitch angle of the spiral arms and the relative
radial force perturbation. The forcing, proportional to the pitch angle, is increasing
from Sa to Sc types in a linear fashion. For our pitch angle of 15.5$^\circ$, the required ratio
for self-consistency is between 5$\%$ and 10$\%$. As shown in P1, the ratio is a function of galactocentric distance R. The 
authors consider strong spirals those in which the ratio is 6$\%$ or more. 

We found that, in order to obtain relative force perturbations in the 5$\%$ to 10$\%$ range, our model 
requires a mass in the spiral pattern of 0.0175 $M_D$, where $M_D$ is the mass of the disk. 
With that choice, our model predicts a peak 
relative force of 6$\%$, and an average value, over R, of 3$\%$. Other spiral masses were explored, but
the analysis favours this case, borderline but on the weak side of the limit        
separating linear (weak) and non-linear (strong) regimes considered by Contopulos \& Grosb{\o}l (1986, 1988).
It is worth noticing that previous results were obtained using rather simplified galactic models,
in which the relative
amplitude of the spiral perturbation was taken as a fixed few percent of the
axisymmetric force at all radii. To illustrate a reference value, Yuan (1969) proposed 5$\%$.

We follow the technique of Contopoulos \& Grosb{\o}l (1986) to compute the ratio of the average
density response and the imposed density, $\rho_{r} /\rho_{i} (R)$, calculating a series of central periodic orbits and
using flux conservation between every two successive orbits. 
A dynamically self-consistent model will be one in which $\rho_{r} /\rho_{i} (R)$ does not deviate 
significantly from unity at any R; it will be a potential in which the orbits of stars will produce
density enhancements in phase with the imposed pattern.  
Hence, $\rho_{r} /\rho_{i} (R)$ is a merit function for the dynamical 
self-consistency of the proposed spiral pattern. We then sample the unknown parameter $\Omega_{p}$ throughout
a fine grid of values, and determine its optimal value as the one for which $\rho_{r} /\rho_{i} (R)$ is as flat
and as close to unity as possible over the range R: [3.3, 12] kpc. This yields $\Omega_p$ = 
20 $km~s^{-1}~kpc^{-1}$ as a prediction of our analysis. The response is quite sensitive to $\Omega_p$:
nearby values such as 19 or 21 $km~s^{-1}~kpc^{-1}$ showed noticeable differences. The range explored
spanned values from 12 to 25 $km~s^{-1}~kpc^{-1}$.

In Figure 1 we show the results for the case
$\Omega_p$ = 20 $km~s^{-1}~kpc^{-1}$. On the Galactic plane, the assumed spiral pattern
and a set of stellar periodic orbits are drawn, with response density maxima  
shown by black squares. Notice the close coincidence  of these with the locus of the imposed pattern
within the boxy-like orbit which marks the 4:1 resonance. In the old kinematic-wave interpretation of
orbital support for the spiral, one can see support inside the resonance, and an abrupt change
corresponding to an off-phase response to the spiral outside it. 
Also dependent on $\Omega_p$ is the fact here displayed that the pattern 
would be dynamically terminated at the position of the 4:1 resonance (here at R = 7 kpc), as
predicted by Contopulos and collaborators. 

\section{Calculating the Gaseous Response to the Best Fit Potential}

Having determined shape, number of arms and total mass content of the spiral pattern
from observations, and having inferred the optimal $\Omega_{p}$ from dynamical self-consistency,
we now study the response of the gaseous disk to such a spiral potential through hydrodynamical 
simulations.

Figure 2 shows the gas response to the imposed pattern with $\Omega_p$ = 20 $km~s^{-1}~kpc^{-1}$.
The locus is indicated with open squares. We performed 2D numerical simulations in polar
coordinates using the ZEUS code (Stone $\& $ Norman, 1992, a,b), without including gas self gravity. The
numerical grid covers 2$\pi $ radians and a radial range from 1 to 15  Kpc; however, we
disregard results internal to R = 3.3 kpc, due to the influence of boundary conditions
(b.c.). The calculation is made in the rotating spiral pattern frame of reference.
Resolution is 500x500 zones and the b. c. are inflow-outflow (inner to outer) in R and
periodic in the azimuth $\phi $. The temperature was fixed at 8000 K, and the
simulation is isothermal given the short cooling timescales compared to the dynamical timescales.
The disk reaches a nearly steady 
state rapidly, which was followed  in this case for 3 Gyr. 
The system is initialized with velocities from the Galactic model rotation curve,
adding the spiral source terms through an input table for ZEUS, and an exponential
gas density law with a radial scale length of 15 kpc and a local
value of about 1.1 $cm^{-3}$ (see Martos \& Cox 1998). 

In between the imposed two-armed pattern, another two-armed
pattern emerges, which displays a sharp shock with a maximum strength along the spiral
between, roughly, 5 and 7 kpc in R. 
At  R around 5 kpc, the arm bifurcates, and the shock 
strength weakens near R = 7 kpc. This position is quite close to the 4:1 resonance. On
each side, this new ``optical" two-armed pattern ends up in a corotation island. In the
figure, the Solar position is along a radial line from the Galactic center (the origin of
both the inertial and the rotating frames) at 20$^\circ$ from the primed (rotating) \'x axis  
(Freudenreich 1998). Following the ``K band" pattern there is a slightly offset gaseous response to the
imposed potential, making the gas response a four-armed pattern. 

A number of caveats apply to our gas simulation: one is
that the strength of the shocks will considerably diminish in a full 3D, MHD simulation
(the inclusion of the vertical direction and magnetic field was considered in Martos \& Cox 1998).
Recent  simulations in full, realistic Galactic models are
scarce: G\'omez \& Cox (2002) employed the Galactic model of Dehnen and Binney (1998).
However, their value of $\Omega_p$ (12 $km~s^{-1}~kpc^{-1}$) places the 4:1 resonance
beyond 22 kpc, far out from the observed pattern extent.  For that $\Omega_p$,
bifurcations and much of the rich
structure is removed and the pattern becomes ring-like (Y\'a\~nez \& Martos 2004). Another important 
component not included yet in these simulations is the Galactic bar, which we recently modeled 
(Pichardo, Martos \& Moreno 2004, hereafter P2). Englmaier \& Gerhard (1999) studied the gas dynamics
in the bar potential determined by data. Their value of $\Omega_p$ is 60 $km~s^{-1}~kpc^{-1}$.
This work reproduces many features of the inner Galaxy. There is then
a large difference in the angular speeds of bar and spiral. The bar-spiral coupling was considered by 
Bissantz, Englmeier \& Gerhard (2003) in a comprehensive study that assembles the available data sets
and SPH simulations in gravitational potentials determined from the near-infrared luminosity of
the bulge and disk, and in some cases, an outer halo and a spiral model for the disc. This work
contains models allowing for different pattern speeds for bar and spiral. Their optimal values,
at the same position of the bar corotation and bar orientation adopted here,
are found to be 60 and 20 $km~s^{-1}~kpc^{-1}$ for the bar and
the spiral, respectively. 

For the spiral pattern, Drimmel \& Spergel (2001) find that the arm strength begins to fall off at about
.85$R_{Sun}$. This is just too close to the R = 7 kpc position of the 4:1 resonance (for our determined
$\Omega_p$)
to dismiss it as an accident.  It seems to
favor a scenario in which the details of the coupling bar-spiral are not crucial to the
dynamics at radii greater than 7 kpc or so. P2 proposes
three different models for the Galactic bar mass distribution: an inhomogeneous
ellipsoid, a prolate spheroid, and a superposition of ellipsoids fitting the observed boxy isophotes.
There are strong model-dependent kinematics near the bar.
Follow up work indicates to us that, while there is a large dispersive effect on orbits inside or near the bar 
region, at the Solar position the effect is minor. For instance, bar-induced vertical dispersive motion occurs
only inside R=4 kpc. However, see M\"{u}hlbauer and Dehnen (2003), who conclude that at 
least the lowest order deviations from
axisymmetric equilibrium in the local kinematics can be attributed to the bar. The comparison with
the effects of the spiral structure is deferred to a future paper. We argue that the result obtained     
by Bissantz, Englmeier \& Gerhard (2003), in a study incorporating dynamics of bar and spiral,
is a strong support to our claim that, even without the presence of the bar
in our model, our predicted value should provide a valid comparison to first order with independent studies   
relying on $\Omega_p$ at the Solar position and beyond.

\section{Recent independent determinations of $\Omega_p$}

From the data in Figure 2 one can directly obtain the gas surface density along a circle with
the radius of the Sun's orbit. In the most simplistic circular approximation (the orbit has
radial excursions of the order of 2 kpc), there are two main peaks of similar densities,
and several local maxima of lower density. The mass contrast is consistent with K-band 
observations (Kranz, Slyz \& Rix 2001),
who quote an arm/interarm density contrast for the old stellar population of 1.8 to
3 for a sample of spiral galaxies. Interestingly, these two peaks have surface densities in 
reasonable agreement with the expected threshold density for star
formation, a value of approximately 10 M$\odot~pc^{-2}$ (Kennicutt 1989),
in which we are considering the reduction in the shock compression due to the magnetic field
and 3D dynamical effects. Other local maxima are factors of 3 or more below that value,
making any associated burst of star formation a less likely event. 

With $\Omega_p$ = 20 $~km~s^{-1}~kpc^{-1}$, the time baseline in the rotating frame for
a circular orbit of that radius is very approximately 1 Gyr
at our assumed Solar radius. We found that the two density peaks mentioned above are separated
in time by $\approx $ 0.5 Gyr. We expect then that periodicity for enhanced star formation.
We note that if self gravity were included the density contrast would increase,
making the predicted periodicity for star formation more evident.      

It is interesting to compare this prediction with recent results of direct inferences
of the star formation history of the solar vicinity. Hernandez, Valls-Gabaud and Gilmore (2000)
analyze the colour-magnitude diagram of the solar neighbourhood as seen by the Hipparcos
satellite, and using Bayesian analysis techniques derive the star formation history over the last
3 Gyr. These authors find an oscillatory component with a period of 0.5 $\pm0.1$ Gyr. By studying 
the age distribution of young globular clusters, de la Fuente Marcos \& de la Fuente Marcos (2004) obtain
a periodicity in the recent star formation history at the solar circle of 0.4 $\pm0.1$ Gyr.
It is remarkable that two distinct and independent assessments of the recent star formation history
at the solar circle yield a periodicity which is perfectly consistent with the density arm crossing
period we derive in this study.

Shaviv (2002) finds that the cosmic ray flux reaching our Solar System should periodically increase
with each crossing of a Galactic spiral arm. Along the same lines of last section, 
over a time baseline of the past 1 Gyr, we
added our estimated magnetic field compression from Martos \& Cox (1998) to plot the expected
synchrotron flux variations moving along the Solar circle and assuming the mass distribution fixed in time.
We find 6 local maxima with fluxes that are higher than today's flux.
This is the same number of peaks satisfying that condition in
Shaviv's work, who plots the ratio cosmic ray flux/today's cosmic ray flux obtained from a
sample of 42 meteorites, which they relate to climate changes in Earth.  
Shaviv (2002) reports a period of 143 Myr for the episodes (crossings) from meteorites data,
which leads to  a value of $\mid \Omega_p$ - $\Omega_{Sun}\mid$ $\ga$ 9.1 $\pm$
2.4 $km~s^{-1}~kpc^{-1}$, which is marginally consistent with our preferred value.       

Andrievsky et al. (2004) report on the spectroscopic investigation of 12 Cepheids situated at
Galactic radii  of 9 to 12 kpc, where they find an abrupt change in metallicity.
The region 10 to 11 kpc appears to be the most important, and the change in metallicity is
explained in terms of the assumed position of corotation. That position is precisely            
the location of corotation in our modeling, if the value $\Omega_p$ =
20 $~km~s^{-1}~kpc^{-1}$ is adopted. However, it is worth noticing that the assumed $R_{Sun}$
is 7.9 kpc, 0.6 kpc less than the value for this fundamental parameter in our Galactic model. This
difference might not alter their results for the change in metallicity level in the vicinity
of corotation, given the large width of the density enhancement caused by the corotation islands 
at about R = 11 kpc in two extended regions of our Figure 2.

\section{Discussion}

As found by Contopoulos \& Grosb{\o}l (1988), self-consistency is improved
by introducing velocity dispersion; this is a realistic effect that can only be neglected by arguing
that in strong spirals nonlinearity dominates. For our Galaxy, the observations of Drimmel \& Spergel
(2001) suggesting the termination of the spiral at the position of the 4:1 resonance indicate       
a marginally strong spiral in Contopoulos \& Grosb{\o}l's (1988) framework. On the opposite side,
our best self-consistent model is found at the lowest spiral mass considered in P1, 0.0175 $M_D$,
for which no stochastic motion was found. From this fact, a weak spiral and a linear regime come to mind.
A possible solution to this issue could be that, while the strength of the spiral begins to fall at the 4:1 resonance,
termination occurs at corotation, as predicted for the weak case in that framework. However, the
quite different modeling of the galactic -- particularly, spiral -- mass distribution used in
their studies and ours could make an interpretation of our results in their framework an unfair one.
At any rate, the response at our best $\Omega_p$ is so flat that there appears no need to
invoke velocity dispersions. In an analysis based on periodic orbits, such as ours, such dispersion
will be small. Other close values of $\Omega_p$ could be improved by this effect in a study out of
the scope of this work, involving many orbits departing from the periodic ones and hence subjected
to larger velocity dispersions.   

We conclude that a two-armed density spiral imposed taking into account observational restrictions from 
K-dwarf distributions yields an optimal dynamically self-consistent model, for values close to
$\Omega_p$ = 20 $~km~s^{-1}~kpc^{-1}$. The fact that various independent estimates of this quantity,
sensitive to highly distinct physics, yield values for
$\Omega_p$ in agreement with our estimation,
gives us confidence in the result. In regard to the gaseous response, we notice that the independent studies
we quote for comparison are not only consistent with the value found here for $\Omega_p$, but also
with a density distribution corresponding to a 4-armed gaseous pattern with structural features
reminiscent of optical observations of our Galaxy and other spirals.

\begin{figure}
\psfig{file=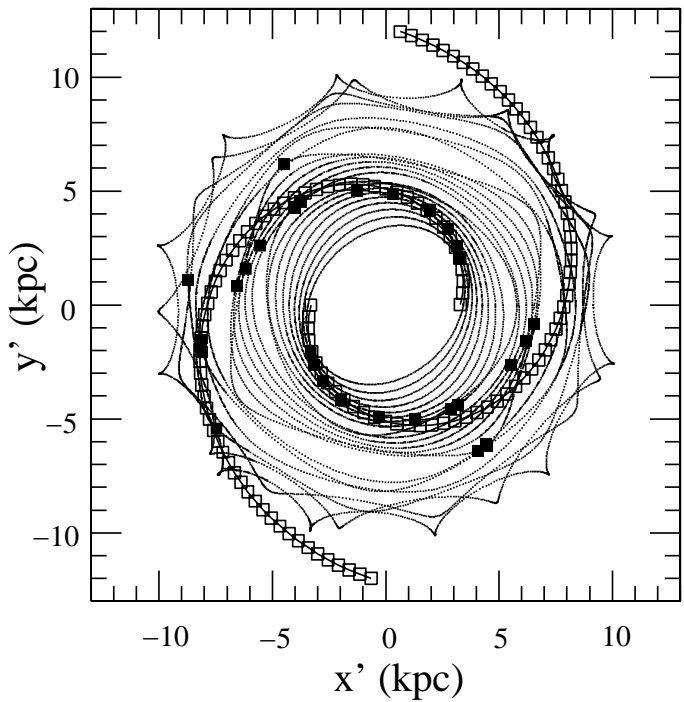,angle=0,width=9.0cm}
\caption{Self-consistency analysis for $\Omega_p = 20 ~km~s^{-1}~kpc^{-1}$. The proposed
spiral locus is shown with open squares. A set of periodic orbits are traced with continuous
lines, and the maxima in the response density are the filled (black) squares. The frame of reference 
is the rotating one where the spiral pattern is at rest} 
\label{fig1}
\end{figure}

\begin{figure}
\psfig{file=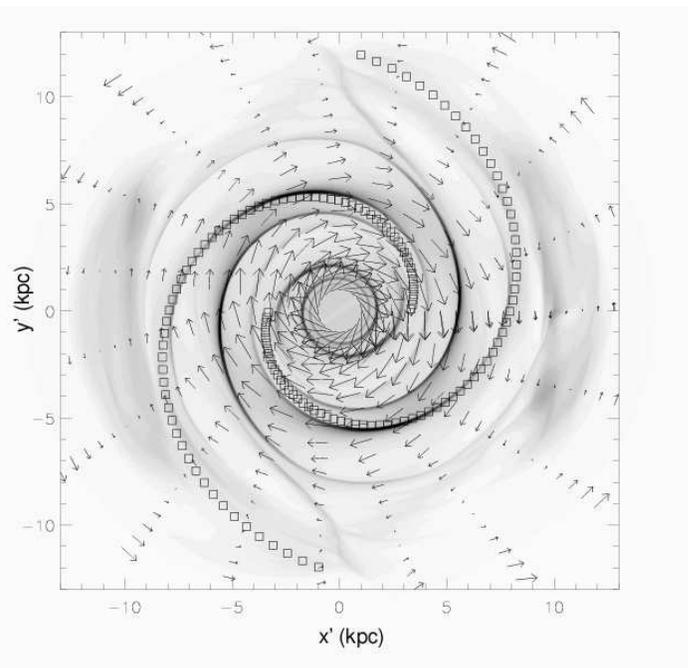,angle=0,width=9.0cm}
\caption{Simulation with the ZEUS code of the gas response to a spiral pattern with 
$\Omega_p = 20 ~km~s^{-1}~kpc^{-1}$, open squares, shown in the
rotating frame of the spiral pattern after 2.55 Gyr of evolution.  
The arrows give the velocity field, their size being proportional to the speed,
with the maximum speed shown being 212 $km~s^{-1}$.
Dense zones correspond to dark regions.}
\label{fig2}
\end{figure}

\section{ACKNOWLEDGMENTS}
We thank an anonymous referee for helpful comments which significantly improved the clarity of the final 
presentation. 
E. Moreno, M. Martos, B. Pichardo, M. Y\'a\~nez acknowledge financial support from 
{\bf CoNaCyT} grant {\tt 36566-E}, and UNAM-DGAPA grant IN114001.

\end{document}